\documentclass[preprintnumbers,prb,floatfix]{revtex4} 
\usepackage{amssymb,amsmath}
\usepackage{color}
\usepackage{graphicx}
\usepackage{dcolumn}
\usepackage{bm}
\usepackage{upgreek}
\usepackage{marvosym }
\usepackage{latexsym,epsfig}
\usepackage{hyperref}
\usepackage{url}

\begin{document}
\title{Two dimensional electron gas in the $\delta$-doped iridates with strong spin-orbit coupling: La$_\delta$Sr$_2$IrO$_4$} 
\author{Churna Bhandari and S. Satpathy}
\affiliation{Department of Physics \& Astronomy, University of Columbia Missouri,
Columbia, MO 65211}
\begin{abstract}

Iridates are of considerable current interest because of the strong spin-orbit coupling that leads to a variety of
new phenomena. 
Using density-functional studies, we predict the formation of a spin-orbital entangled two dimensional electron gas (2DEG)
in the $\delta$-doped iridate La$_\delta$Sr$_2$IrO$_4$, where a single SrO layer is replaced by a LaO layer. 
The extra La electron resides close to the $\delta$-doped layer, partially occupying the  $J_{\rm eff}= 1/2 $ upper Hubbard band and
thereby making the interface metallic. The magnetic structure of the bulk  is destroyed near the interface, with the
Ir$_0$ layer closest to the interface becoming non-magnetic, while the next layer (Ir$_1$) continues to maintain the  AFM structure of the bulk, but with a reduced magnetic moment. 
The Fermi surface consists of a hole pocket and an electron pocket, 
located in two different Ir layers (Ir$_0$ and Ir$_1$), with both carriers
derived from the  $J_{\rm eff}= 1/2 $ upper Hubbard band.
The presence of both electrons and holes at the $\delta$-doped interface suggests unusual transport properties,
leading to possible device applications.

\end{abstract}

\date{\today}						

\maketitle

The combination of a large spin-orbit coupling (SOC) and a reduced Coulomb interaction in the 4d and 5d transition metal oxides 
has made these compounds hosts for a number of exotic quantum states, such as the spin-orbit driven Mott insulators\cite{KimPRL08},  
Weyl semimetals\cite{YuPRB11}, axion insulators\cite{KrempaARCMP,WanPRL}, and Kitaev spin liquids.\cite{Balentsnat10,LawlerPRL08} 
Analogously, one expects any two dimensional electron gas (2DEG) formed in the Sr$_2$IrO$_4$ (SIO) iridate structures\cite{BhandariPRB18}, similar to the 3d interfaces
such as LaAlO$_3$/SrTiO$_3$\cite{PopovicPRL08}, to lead to potentially novel effects due to the spin-orbital entanglement of the electron states. 
Recently, epitaxial growth of iridate heterostructures and interfaces has been demonstrated by a number of 
experiments\cite{MatsunoPRL15,  MingCPL15, Groenendijk16}, which paves the way for the engineering of 
$\delta$-doped iridates structures by substitution of an element with another having an excess electron or hole. 
Such $\delta$-doped structures
can further lead to the formation of 2DEG near the dopants.
 

The characteristic electronic properties of the iridates are controlled by the Ir$^{4+}$ ion in the d$^5$ configuration. A large crystal field splits the d 
states into two e$_g$ and three t$_{2g}$ states. The sixfold t$_{2g}$ manifold (including spin) is further split into a fourfold $J_{eff} = 3/2$ and a 
twofold $J_{eff}= 1/2$ manifold, with an energy gap of $\lambda_{SO} = 3 \lambda /2$,
$\lambda \vec L \cdot \vec S$ being the SOC interaction. Four electrons fill the lower-lying $J_{eff} = 3/2$ states, leaving the lone d electron to occupy
 the doubly degenerate $J_{eff} = 1/2$ state. 
The wave functions $| J_{eff}, m \rangle$  are spin-orbital entangled, viz., 
\begin{eqnarray}
  |e_1\rangle &\equiv |\frac{1}{2}, -\frac{1}{2} \rangle =
(|xy\uparrow\rangle + |yz\downarrow\rangle + i|xz\downarrow\rangle)/\sqrt 3,\nonumber  \\
|e_2\rangle &\equiv |\frac{1}{2}, \frac{1}{2} \rangle =
(|yz\uparrow\rangle - i|xz\uparrow\rangle - |xy\downarrow\rangle)/\sqrt 3.
\label{eq1}
\end{eqnarray}
The half-filled $J_{eff} = 1/2$ band splits further due to the Coulomb interaction into an upper Hubbard band (UHB) and a lower Hubbard band (LHB), producing a Mott insulator. 

The Mott insulating state can be doped with electrons or holes via impurity substitution. Experimentally, there have been already several studies on the bulk electron doping \cite{MingCPL15,GePRB11,KornetaPRB12,NicholasPRL13,SerraoPRB13,MiaoPRB14,LeePRB12,KornetaPRB10,KimSci14} 
by the substitution of Sr by trivalent metals like La.\cite{GretarsonPRL16,ChenPRB15} These bulk doped samples display 
several interesting features such as the formation of Fermi arcs and  pseudogaps by potassium surface doping\cite{KimSci14} and a metallic 
state beyond $5\%$ \cite{TorrePRL15} or $15\%$ La.\cite{MingCPL15}

In this paper, we predict the formation of a 2DEG in the $\delta$-doped SIO iridate 
using density functional methods. The predicted 2DEG is different from a conventional 2DEG 
obtained at the $\delta$-doped  3d perovskite oxides such as La$_\delta$$\rm SrTiO_3$ structures\cite{Popovic05, MatsubaraNat16} in several ways,
viz., (i)  The 2DEG is spin-orbital entangled,  
(ii) It
is sharply localized on just two Ir  (Ir$_0$ and Ir$_1$) planes adjacent to $\delta$-doped LaO layer 
as opposed to the La$_\delta$$\rm SrTiO_3$, 
where it spreads to several  unit-cell layers around LaO, 
and (iii) As opposed to the multi-band Fermi surface for the 3d structures, the Fermi surface is much simpler in the present case,
consisting of just one (nearly circular) hole  pocket at $\Gamma$ and an elliptical electron pocket at $M$ point of the 2D interface Brillouin zone.


\begin{figure}[tbp!]
\includegraphics[scale=0.6]{cor} \includegraphics[scale=0.6]{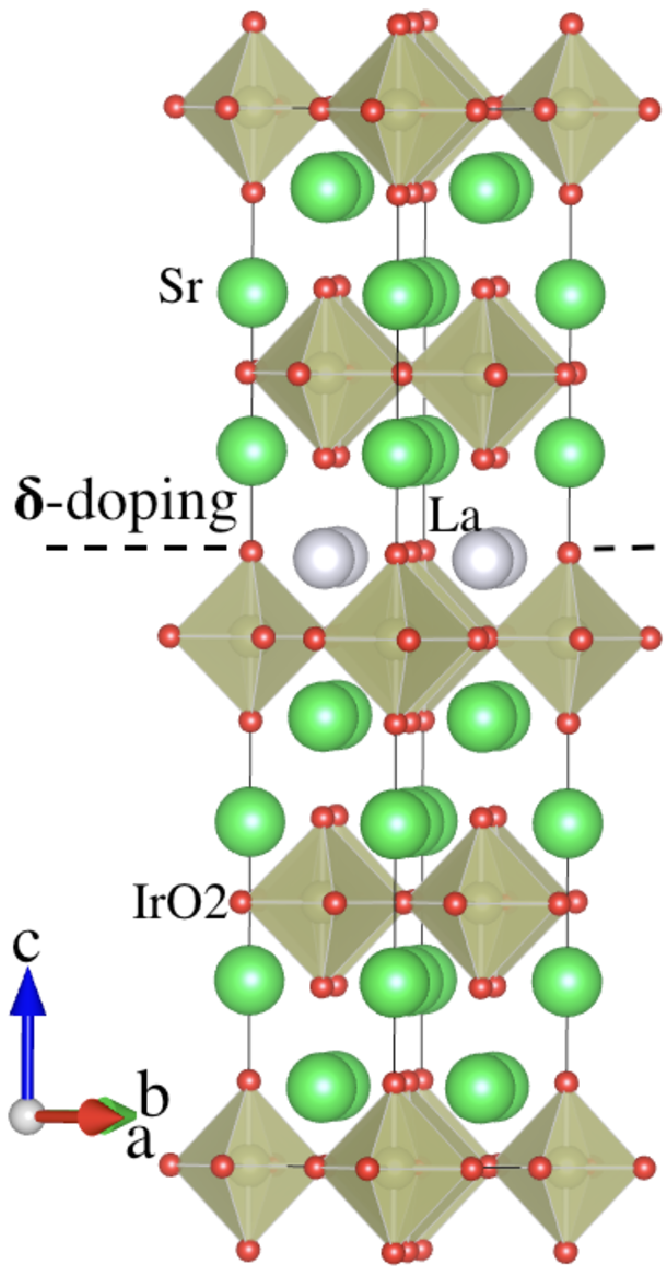}\\
\hspace{ 1 cm}   (b)      \hspace{4.5cm}     (a)\\
\includegraphics[scale=0.35]{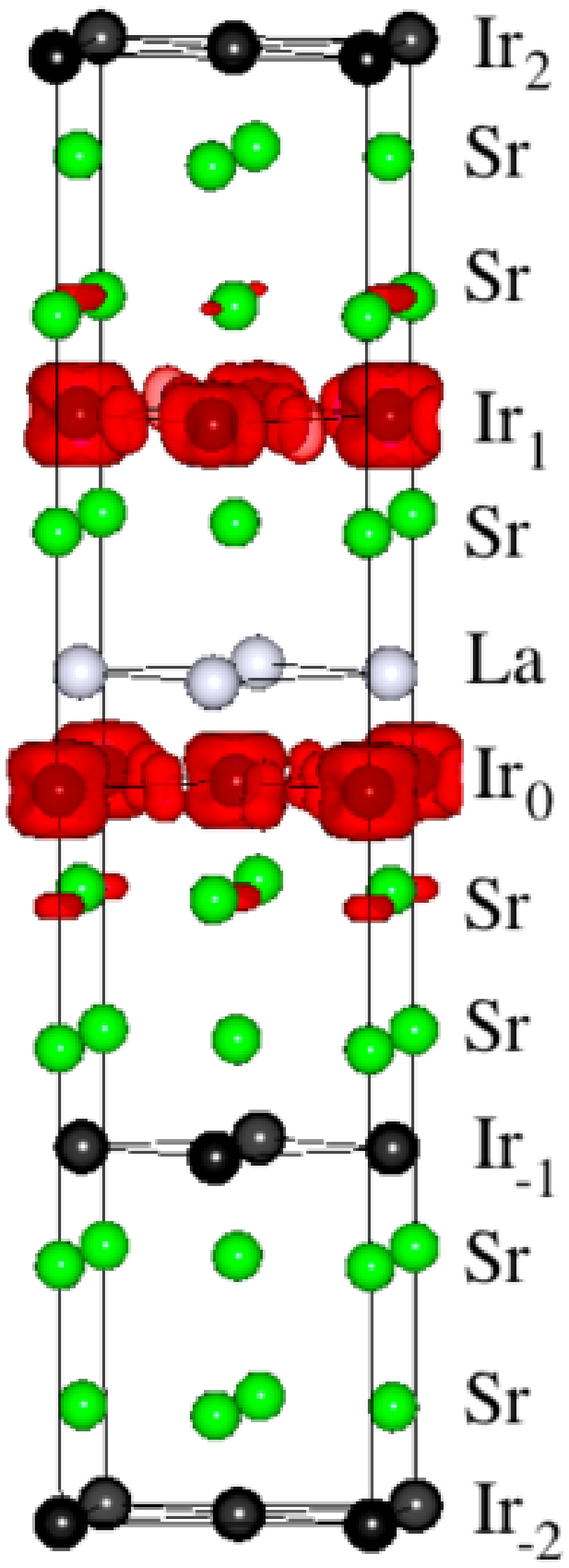}\\
                (c)
\caption{(a) The $\delta$-doped La$_\delta$Sr$_2$IrO$_4$ structure, with a portion of the 
supercell shown, (b) The cell averaged O $1s$ core level energies as a function
of the distance from the LaO layer, showing a V-shaped  potential, and (c) Charge density contours for the 
electron bands near $E_F$, indicating the localization of the 2DEG on the two Ir layers closest to LaO.
A small slice of energy (in the range of $E_F$ and $E_F$ - 0.15 eV) was used to compute the charge density. The isolevel
shown in red corresponds to the charge density of $10^{-3}$  e/\AA$^{3}$. 
Note that in Fig. (a), the Ir atoms occur at the center of the oxygen octahedra.}
\label{rho}
\end{figure}

In our study, we used  the full-potential linearized muffin-tin orbital  (FP-LMTO) method \cite{Methfessel,Kotani10,lmsuite}  
to solve the density functional Kohn-Sham equations in the 
local spin density approximation\cite{Kohn-Sham,vonBarthHedin} including 
spin-orbit coupling  and the Hubbard U term (LSDA+SOC+U). 
The supercell consisted of 8 SIO layers stacked along the $c$-axis and each layer consisted of two formula units,
so as to describe the possible anti-ferromagnetic state in each layer, leading to the unit cell consisting of (Sr$_2$IrO$_4$)$_{16}$. 
All atom positions were optimized  with the lattice constant of the 
supercell structure  fixed at the bulk experimental value.
To form  the $\delta$-doped structure, we substituted a single layer of SrO by LaO,
as indicated in Fig. \ref{rho}.
We also used the Vienna {\it ab initio} Simulation Package (VASP)\cite{KressePRB93} within the projector augmented wave (PAW) \cite{BlochlPRB94} method including $U$ in the Dubarev 
scheme\cite{Dubarev} and SOC for computing magnetic moments. We used the plane wave cut-off energy of $620$ eV and a
$6\times6\times1$ $ {\bf k}$-mesh for Brillouin zone sampling.


On general grounds, one expects a $V$-shaped potential at the $\delta$-doped layer.
Since La has an extra valence electron as compared to Sr, 
with the transfer of these electrons to the solid, the La$^{+3}$O$^{-2}$ layer becomes nominally a positively charged layer 
as compared to the neutral Sr$^{+2}$O$^{-2}$ layer, and thus it produces a uniform electric field on either side of the infinite 2D plane, leading to a $V$-shaped attractive potential, in which the extra valence electron of La becomes bound.

The potential due to a charged sheet is simply $V= (\sigma/\varepsilon ) |z|$, where $\sigma$  is the surface charge density,
$\varepsilon$ is the dielectric constant, and $z$ is the distance from the interface.
The potential near the $\delta$-doped layer may be examined by studying the core level energy shifts.
Using density functional theory (DFT), we have
computed the O ($1s$) core energies. The cell averaged  core level energies  (averaged over the bulk unit cell)
are shown in Fig. \ref{rho} (b). 
As one moves away from the interface, the interfacial potential is screened due to the redistribution of electrons around the donor plane. 
From the DFT calculations, we find a more or less 
$V$-shaped potential close to the interface. 
Apart from this, Fig. \ref{rho} (b) shows that the Ir layers
as close as the second layer to  LaO (viz., Ir$_2$ and Ir$_{-1}$)  already look like the bulk.

We find that the interface potential is strong enough to localize the $\delta$-doped electrons in the two Ir layers closest
to the interface. This may be seen from the charge density of the electron bands near the Fermi energy ($E_F$).
For this purpose, we used the energy range of $E_F$ to 0.15 eV below it. 
The results shown in Fig. \ref{rho} (c) indicate the confinement of the doped electrons to mostly the Ir$_0$ and Ir$_1$ layers.
The same conclusion is inferred from the computed charges of the individual atoms
shown in Table \ref{tab1}, 
which were estimated by computing the total charges within the muffin-tin spheres and renormalizing them to the
total number of electrons in the system.

\begin{table}
\caption{Distribution of the $\delta$-doped electron (one per La atom)  
 and the  
spin and orbital magnetic moments (in units of $\mu_{B}$)  of the Ir atoms in the layers close to the interface. 
The listed charges are in essence the occupancy $n_2$ of the UHB  belonging to different Ir layers, 
while the corresponding occupation of the  LHB is $n_1 = 1$ for each Ir layer.
}
\begin{ruledtabular}
\begin{tabular}{l|cc|cc|cc}
atom &Ir$_{0}$ & Ir$_1$ &Ir$_2$ &Ir$_{-1}$ & rest\\
\hline
charge (e) & 0.64 & 0.18 & 0 & 0 &0.18\\
$\mu_s $ & 0.00 & 0.06 & 0.11 & 0.11 & 0\\
$\mu_l $ & 0.00 & 0.14 & 0.28 & 0.28 & 0\\
\end{tabular}
\end{ruledtabular}\label{tab1}
\end{table}

\begin{figure} [tbp!]
\includegraphics[scale=0.50]{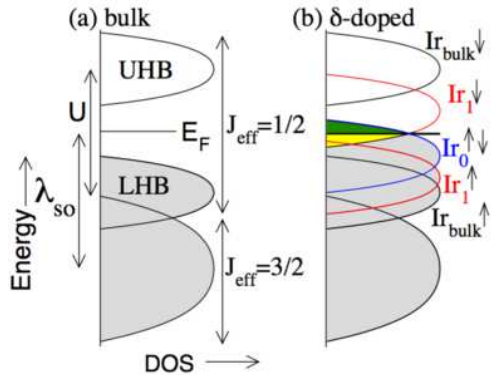}
\caption{ Schematic electronic structure for the Ir t$_{2g}$ bands for bulk SIO (a) and the $\delta$-doped structure (b), as obtained from the 
DFT calculations. 
}
\label{schematic}
\end{figure}

\begin{figure} [tbp!]
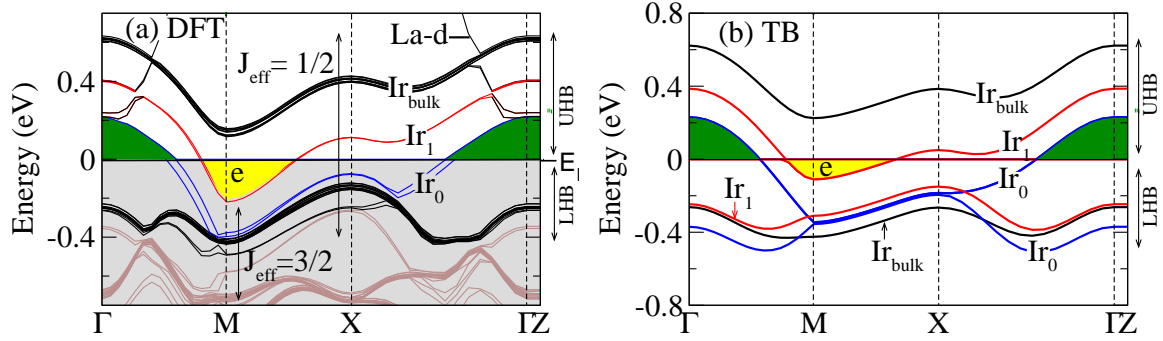

\vspace{4cm}
\includegraphics[scale=0.45]{band.eps}
\includegraphics[scale=0.45]{tb.eps}
\caption{
Band structure of La$_{\delta}$Sr$_2$IrO$_4$ computed using DFT (a) and the tight-binding model (b). 
The 
yellow region represents the electron pocket in the UHB for
Ir$_1$, while the green region corresponds to the hole pocket in Ir$_0$. 
Both the UHBs and the LHBs originating from various Ir layers are color coded in the TB results, Fig. (b), while in the DFT bands, Fig. (a),
only the UHBs can be clearly identified as shown.
}
\label{fig3}
\end{figure}

\begin{figure} [tbp!]
\vspace*{2cm}
\includegraphics[scale=0.35]{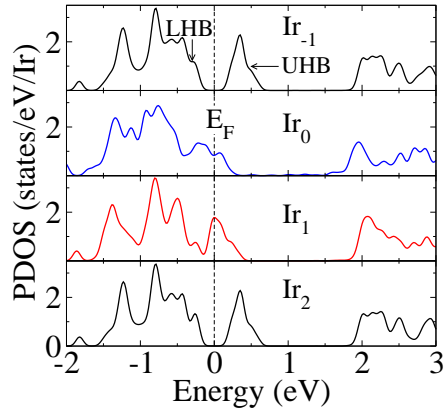}
\caption{
Layer resolved Ir (d) density of states, indicating the empty UHB in the bulk-like Ir$_2$ and Ir$_{-1}$ layers, while the same in the Ir$_0$ and Ir$_{1}$ layer are partially filled to accommodate the $\delta$-doped electrons.}
\label{fig4}
\end{figure}

\begin{figure} [tbp!]
\includegraphics[scale=0.475]{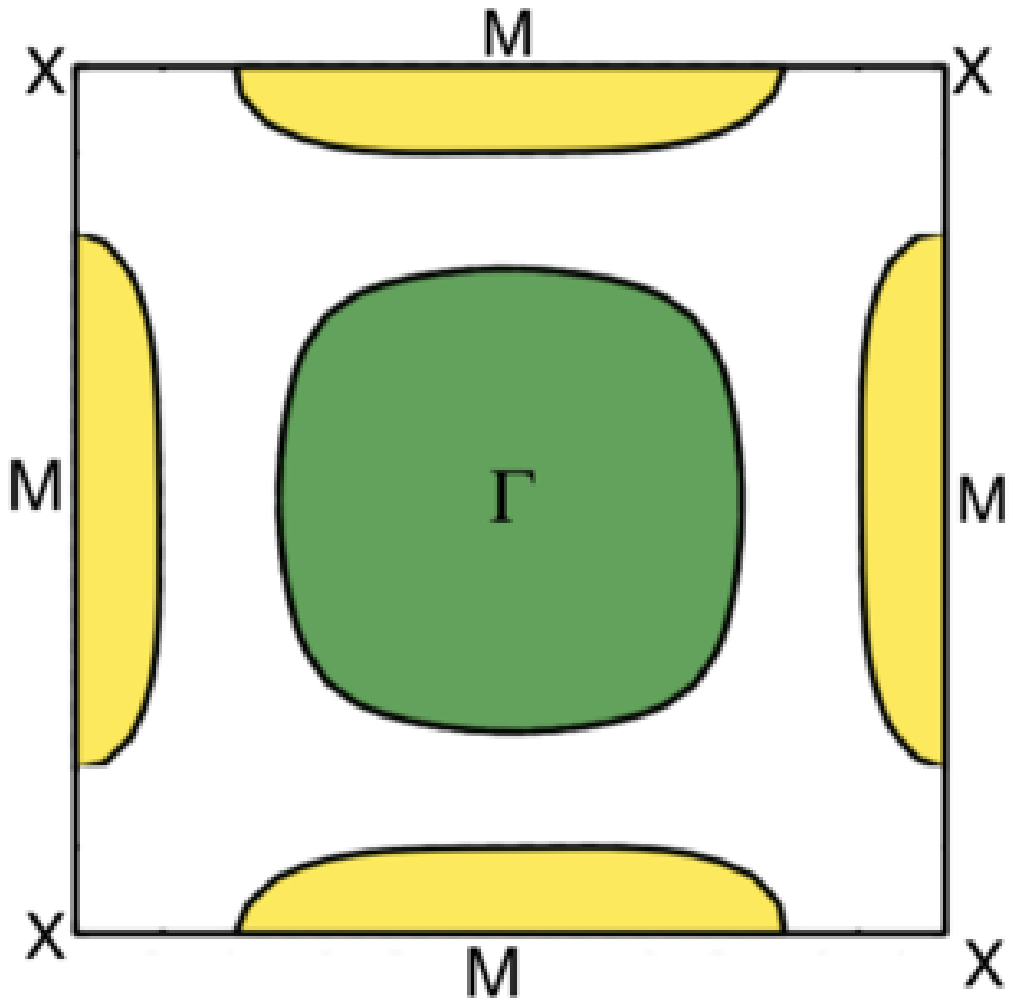} 
\caption{
Fermi surface of the $\delta$-doped structure, consisting of a hole pocket at $\Gamma$ (green)  and an electron pocket at $M$ (yellow). }
\label{fig5}
\end{figure}

We now turn to the band structure and the Fermi surface. Fig. \ref{schematic}  (a) shows the well known\cite{KimPRL08}  spin-orbit assisted Mott insulating state of the bulk SIO, in terms of which the electron states of the $\delta$-doped structure, shown in 
Fig. \ref{schematic}  (b), may be understood. The individual Ir layers in the structure are separated enough that
the electronic structure can be understood in terms of the isolated Iayers. 
The upper and the lower Hubbard bands for the interface Ir layers (Ir$_0$ and Ir$_1$) are shifted in energy due to the V-shaped interface potential and, in addition, they show a layer-dependent splitting $ 2\Delta = U (n_1 - n_2)$ between the 
LHB and the UHB, with the respective population being $n_1$ and $n_2$.
The LHB is full for each Ir layer ($n_1 = 1$), while the UHB is only
partially full, $n_2 < 1$ for Ir$_0$ and Ir$_1$, and 0 for the remaining Ir atoms, which are similar to the SIO bulk.
The charges listed in Table \ref{tab1} correspond to the charges $n_2$ in the UHB for the various layers.

Fig. \ref{fig3}  shows the density functional band structure, 
computed using the FP-LMTO method and the 
LSDA+SOC+U functional ($U= 2.7$ eV was used following earlier works on bulk SIO), 
as well as the model tight-binding (TB) results.
The Ir (d) $J_{eff} = 1/2$ bands, split into the UHB and LHB, are the ones that occur around the Fermi energy, and the 
UHB belonging to different Ir layers can be identified in the band structure as marked in Fig. \ref{fig3} (a). The layer-projected DOS shown in Fig. \ref{fig4} indicates the partial occupation of the 
UHB of the Ir atoms close to the $\delta$-doped layer in order to accommodate the doped electrons,
which is consistent with the charge distribution presented in Table \ref{tab1}.

To help in the understanding of the band structure, we have considered the Hubbard model on a square lattice 
with anti-ferromagnetic order as appropriate for a single Ir layer in SIO, keeping the two spin-orbital entangled 
$J_{eff}=1/2$ orbitals ($|e_1 \rangle$ and $|e_2\rangle$ in Eq. \ref{eq1}). 
The TB Hubbard Hamiltonian is given by
\begin{equation}
{\cal H} = \sum_{ \langle ij\rangle   \alpha} t_{ij} c^\dagger_{i\alpha}c_{j\alpha}+h.c. + \frac{U}{2} \sum_{i \alpha}  n_{i\alpha} n_{i \bar \alpha},
\end{equation}
where $c^\dagger_{i\alpha}$ creates an electron at site $i$ (which may be sublattice A or B)   
and orbital $|e_\alpha \rangle$,
 $t_{ij}$ is the hopping integral, $U$ is the on-site Coulomb interaction, and the summation $\langle ij\rangle$
is over distinct pairs of neighbors.  
The same model parameters, which we used earlier\cite{BhandariPRB18,Churna-strain}  to describe the bulk band structure of SIO, were adopted here, viz., 
$U = 0.65$, $t_1 = -0.095$, $t_2 =0.015$, $t_3 = 0.035$, and $t_4 = 0.01$, all in units of eV,
with $t_n$ being the $n$-th nearest neighbor hopping.

In the momentum space, the $4 \times 4$ Hamiltonian (two sublattice sites in the AFM unit cell and two orbitals per site) 
becomes block diagonal with two identical $2 \times 2$ TB Hamiltonian matrices in the Bloch function basis for $| e_1\rangle $ or $|  e_2\rangle $, viz.,  
\begin{equation}
{\cal H} (\bf k) = \left(\begin{array}{cc}  
-\Delta + h_{11}({\bf k})   & h_{12}({\bf k})  \\ 
h_{12}^*(\bf k)  & \Delta + h_{11}(\bf k)    
\end{array}
\right),
\label{H22} 
\end{equation}
where $h_{11}({\bf k})  = 4t_2  \cos k_x \cos k_y + 2 t_3 (\cos 2 k_x + \cos 2 k_y) $, 
$h_{12}({\bf k})  = 2 t_1 (  \cos k_x + \cos k_y)  + 4 t_4  (\cos 2 k_x  \cos  k_y + \cos  2 k_y  \cos   k_x) $, 
 the lattice constant of the square lattice is taken as one, and $\Delta = U/2 \times (n_1 - n_2)$ is the Hartree-Fock staggered field with
 $n_\alpha$ being the occupancy of the $| e_\alpha \rangle$ orbital. 
 One immediately gets the eigenvalues
\begin{equation}     \label{TB-Energy}
\varepsilon_\pm ({\bf k}) = h_{11}  ({\bf k}) \pm \sqrt { \Delta^2 + h_{12}^2  ({\bf k}) },
\end{equation} 
which are plotted in Fig. \ref{fig3} (b), where the TB bands for the individual Ir layers are shifted somewhat to fit with
the DFT bands. 

The bands originating from the various Ir layers are shown in Fig. \ref{fig3} (b), and the UHBs can be clearly identified in the 
DFT bands, Fig. \ref{fig3} (a), as well. Note that the staggered field $\Delta \equiv U (n_1 - n_2)/2$ that characterizes the band splitting between the UHB and the LHB, following Eq. \ref{TB-Energy}, is layer dependent, which leads to different shapes of the bands for different Ir layers, as clearly seen  in the TB band structure.

The band structure clearly indicates that the $\delta$-doped electrons occupy the UHB of the Ir$_0$ and Ir$_1$ layers. 
In addition, the layer resolved Ir densities-of-states (DOS)
shown in Fig. \ref{fig4} confirms the same picture. We estimated the amount of $\delta$-doped charge in each layer by integrating the layer projected DOS, which yields  $64\%$ for Ir$_0$, $18\%$ for  Ir$_{1}$, and the remaining $18\%$ shared among other atoms 
such as La, O etc. These are the charge density values listed in Table \ref{tab1}. 

The Fermi surface, shown in Fig. \ref{fig5}, consists of a hole pocket at $\Gamma$ and an electron pocket at $M$. An interesting point is that the electrons and holes are spatially separated, with the electron pocket occurring in the Ir$_1$ layer, while
 the hole pocket occurs in Ir$_0$.   
 The presence of both an electron and a hole pocket is quite unusual and it happens because of the presence of multiple bands in the electronic structure, with the nearly empty bands behaving as electrons, while the 
 nearly filled bands behaving like holes. We note that this feature is not always found, e. g., in the La $\delta$-doped perovskite La$_\delta$Sr$_{1-\delta}$TiO$_3$, where also a 2DEG is found along with multiple bands, there is only electron-like behavior.\cite{Popovic05}
 Transport measurements should show the presence of these two different types of carriers, unraveling this
 unusual behavior.

The  2DEG alters the magnetic moments of the Ir atoms near the $\delta$-doped layer.
The computed spin and orbital magnetic moments given in Table \ref{tab1} shows that both the spin
and orbital magnetic moments, $\mu_s$ and $\mu_l$ are zero for Ir$_0$ making it non-magnetic,
while for Ir$_1$, they are roughly half of the bulk values, already achieved in the Ir$_2$ and the Ir$_{-1}$ layers.
For these layers, the calculated net magnetic moment is 0.39 $\mu_B$ compared to the calculated
$\approx 0.36$  $\mu_{B}$ for the bulk.\cite{Churna-strain}  
The 
experimental magnetic moment in the bulk has been estimated to be  $0.5 \mu_B$ from  magnetic susceptibility measurements.\cite{CaoPRB98}

In summary,
we predicted the formation of a spin-orbital entangled 2DEG in the $\delta$-doped iridate La$_\delta$Sr$_2$IrO$_4$, where 
a single layer of SrO is replaced by LaO in the Sr$_2$IrO$_4$ crystal. The doped electron remains close to the interface,
which becomes metallic, with an electron and a hole pocket forming the Fermi surface, while the bulk, away from the interface
is insulating. Experimental observation of the 2DEG would be quite interesting as it would result in a novel, spin-orbital entangled electron gas, with properties quite different from the 2DEG in the 3d oxide structures such as the well-studied LaAlO$_3$/SrTiO$_3$ interface.

This research was supported by the U.S. Department of Energy, 
Office of Basic Energy Sciences, Division of Materials Sciences and Engineering under Grant  No. DE-FG02-00ER45818.


\end{document}